\documentclass[10pt]{iopart}

\usepackage{graphicx}
\usepackage{epstopdf}
\usepackage[latin9]{inputenc}
\usepackage{amsbsy}
\usepackage{gensymb}

\begin{document}
\title{Moderate Magnetic Field Induced Large Exchange Bias Effect in Ferrimagnetic 314 - Sr$_{3}$YCo$_{4}$O$_{10.5}$ Material}
\ioptwocol

\author{Sourav Marik, Prachi Mohanty, Deepak Singh and Ravi P. Singh}
\address{Indian Institute of Science Education and Research Bhopal, Bhopal, 462066, India}
\ead{soumarik@gmail.com and rpsingh@iiserb.ac.in}
\vspace{10pt}
\begin{indented}
\item[]September 2017
\end{indented}

\begin{abstract}
Herein, we report the appearance of a large exchange bias (EB) effect in a moderate cooling field (cooling field, H$_{FC}$ = 1 kOe) for the 314 - Sr$_{3}$YCo$_{4}$O$_{10.5}$ material. The exchange bias has started to appear near room temperature and reaches a maximum value of 5.5 kOe at 4 K. The existence of ferrimagnetic clusters in the compensated host in this layered structure originates the large exchange anisotropy. Remarkably, the observed value of moderate magnetic field induced exchange bias field is extremely large in comparison with material systems which are recognized to exhibit giant exchange bias effect. In combination with the feasibility of room temperature application, the appearance of large exchange bias in a moderate cooling field exemplifying the present material system as a promising class of compounds for designing coherent magnetic materials with huge exchange bias in low/moderate magnetic field. 
\end{abstract}
\normalsize
\section{Introduction}
The intriguing fundamental and technological aspects of the exchange bias (EB) effect have garnered great interest since its discovery by Meiklejohn and Bean \cite{1}. Recent advances on EB have highlighted its usefulness in several technological applications such as in magnetic recording read heads \cite{2}, thermally assisted magnetic random access memories \cite{3}, and other spintronic devices \cite{4,5}. Exchange bias is an induced magnetic anisotropy, which occurs in ferromagnetic (FM) materials when coupled to an antiferromagnet (AFM) and yields a shift \cite{1} in the isothermal magnetization loop with respect to horizontal magnetic field axis and vertical magnetization axis. The effect has been therefore intensively studied; however, a comprehensive understanding of exchange bias is a long-standing problem. In general, its phenomenology has variously been interpreted in terms of ferromagnetic (FM)-Antiferromagnetic (AFM) interfaces, that causes a pinning of magnetization such as inhomogeneous magnetic phases like FM/ferrimagnet \cite{6}, FM/spin glass \cite{7}, hard/soft FM systems \cite{8}, magnetic nanoparticles \cite{9}, granular composites \cite{10}, bilayers \cite{11}, and super lattices \cite{12}.
\begin{figure}[h]
\includegraphics[width=1.0\columnwidth]{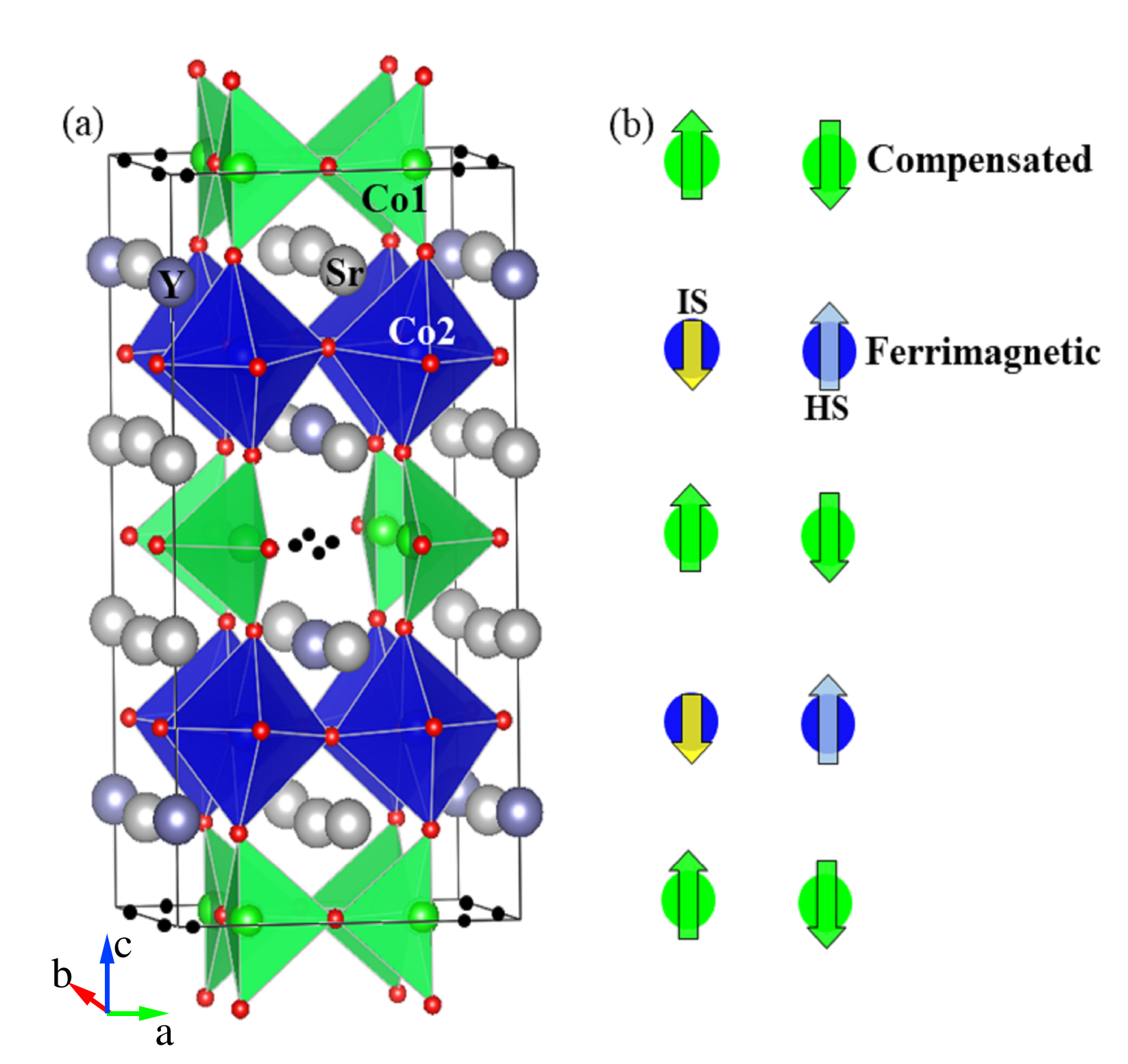}
\caption{\label{Fig1:str} (a) Crystal structure of the oxygen deficient 314 - Sr$_{3}$YCo$_{4}$O$_{10.5}$ material. The black spheres indicate the extra oxygen in comparison with Brownmillerite structure, and only about 1/4 of them are occupied. (b) The magnetic ordering scheme of Co cations along [100] direction for the same.}
\end{figure}

	Recently, nearly compensated ferrimagnetic systems such as Mn$_{3-x}$Pt$_{x}$Ga \cite{13}, Ba$_{2}$Fe$_{1.12}$Os$_{0.88}$O$_{6}$ \cite{14} and SrFe$_{0.15}$Co$_{0.85}$O$_{2.62}$ \cite{15} have attracted immense interest due to their potentiality to show giant exchange bias effect. A small lack of compensation (presence of ferrimagnetic clusters in the compensated host), which is formed as a consequence of intrinsic antisite disorder originates the giant EB effects in these systems and indeed provides a novel approach to design new materials with a large EB. Alongside the EB phenomenon, the compensated materials are proposed to play an important role in antiferromagnetic spintronics \cite{16}.
	
For the feasibility of room temperature (RT) application, it would be worth to explore the exchange bias effect in nearly compensated ferrimagnetic materials having transition higher than room temperature. In this context, the nearly compensated ferrimanetic system Sr$_{3}$YCo$_{4}$O$_{10.5}$ \cite{17,18} with ferrimagnetic transition above room temperature is an interesting system to search for a magnetically compensated material to observe exchange bias. The oxygen deficient 314 material Sr$_{3}$YCo$_{4}$O$_{10.5}$ crystallize in a tetragonal symmetry, space group I4/mmm with a 2a$_{p}$ $\times$ 2a$_{p}$ $\times$ 4a$_{p}$ (a$_{p}$ = lattice parameter of the cubic perovskite sub-cell, Fig. 1) type unit cell and shows the appearance of ferrimagnetism well above RT. In this compound Co$^{3+}$ cations in the oxygen replete layers are antiferromagnetically aligned with alternating high - spin (HS, t$_{2g}$$^{4}$e$_{g}$$^{2}$, S = 2) state and intermediate - spin (IS, t$_{2g}$$^{5}$e$_{g}$$^{1}$, S = 1) state, resulting in above RT ferrimagnetism \cite{17,18,19}.

	In this work, we report the moderate field induced sizable exchange bias effect for the layered 314 - Sr$_{3}$YCo$_{4}$O$_{10.5}$ material. Detailed magnetization measurements are employed to demonstrate the exchange bias effect in this material.
 \section{Experiment}
A pure phase polycrystalline sample of Sr$_{3}$YCo$_{4}$O$_{10.5}$ was prepared by the standard solid-state reaction as reported earlier \cite{17,18}. The sample was characterized by X-ray powder diffraction (XRD) at room temperature (RT), performed in a Panalytical X'pert Pro diffractometer. Temperature and field dependent magnetization measurements were carried out by using Quantum Design PPMS/MPMS - 3 magnetometer.
\begin{figure}[h]
\includegraphics[width=1.0\columnwidth]{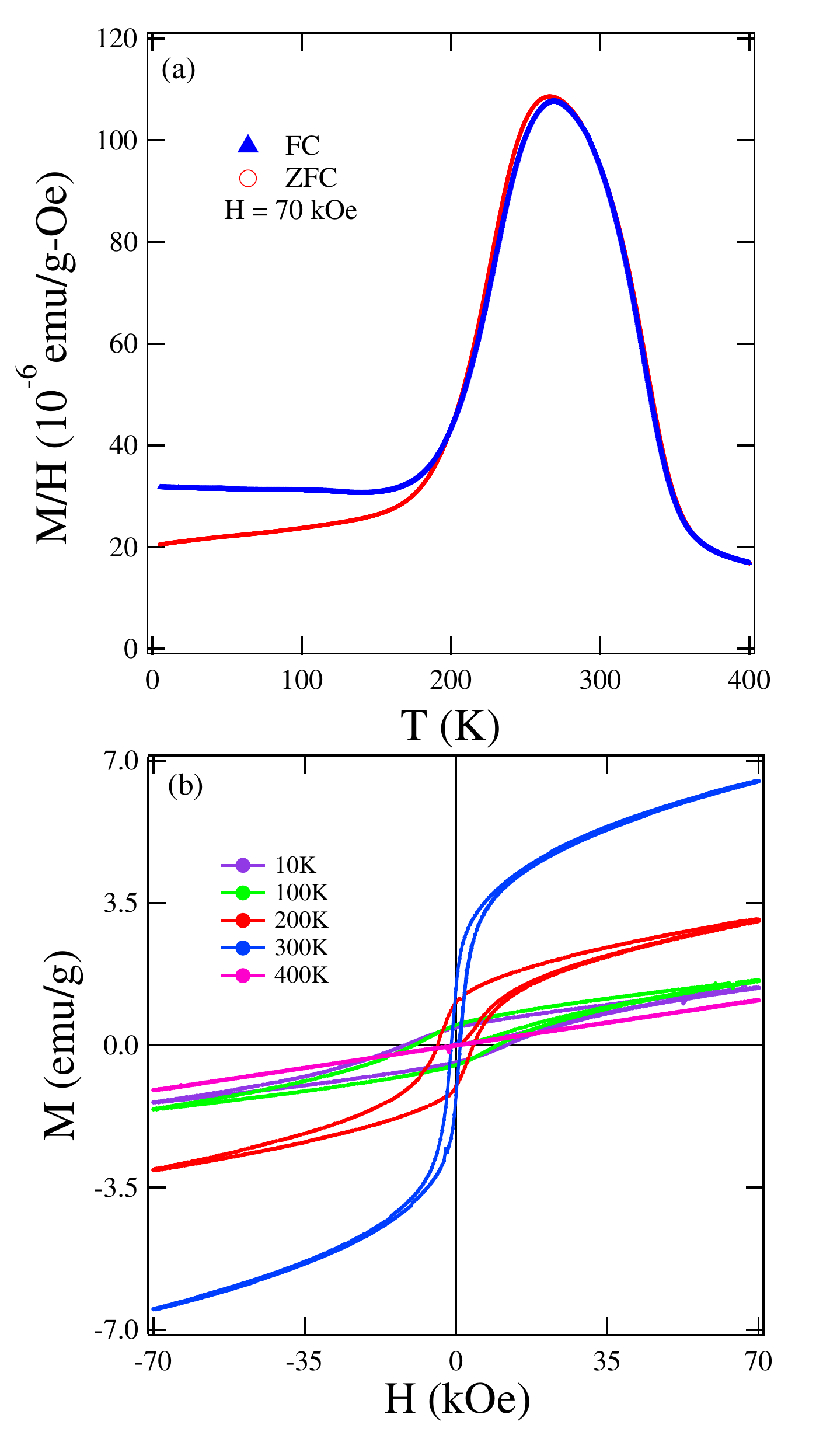}
\caption{\label{Fig2:ds} (a) Temperature variation of magnetic susceptibility. (b) Zero-field-cooled magnetic hysteresis measurements (M-H) at several temperatures for the Sr$_{3}$YCo$_{4}$O$_{10.5}$ material.}
\end{figure}
\section{Results and discussion}

	The RT XRD pattern indicates that the compound can be isolated as a single phase, crystallizing in a tetragonal symmetry with $\textit{I4/mmm}$ space group (S.G.). As suggested previously \cite{17,18,19}, the present structure consists of alternating oxygen-replete and oxygen deficient layers [Fig.1] in close similarity to the Brownmillerite structure \cite{20}. Transmission electron micrograph shows that the particles possess a quasi-spherical morphology with an average size of 0.3 $\mu$m. Fig. 2(a) shows the temperature variation of the field cooled (FC) susceptibility (measured with 70 kOe magnetic field), which confirms the ferrimagnetic transition at 345 K. The ferrimagnetic structure is formed with a Co$^{3+}$ HS state and IS state ordering (AFM aligned) in the oxygen-replete layers, while the Co cations in the oxygen deficient site are nearly compensated (Fig. 1) \cite{18}. The isothermal magnetization measurements performed in zero field cooling mode (ZFC) is shown in Fig. 2(b). The increase in the coercive field at a lower temperature is due to the existence of uncompensated spins (AFM ordered) in a compensated matrix. 
	
	The magnetic hysteresis measurement (M-H) performed at 10 K in field cooled (FC) method with a moderate cooling field, H$_{FC}$ = 1 kOe for the Sr$_{3}$YCo$_{4}$O$_{10.5}$ sample is shown in Fig. 3. For the FC M-H measurement the sample was cooled from T = 400 K $>$ T$_{C}$ $\approx$ 345 K. A clear shift in the FC M-H loop towards both the negative field and the positive magnetization direction is apparent in Fig. 3. This shift can be characterized in terms of exchange bias field (H$_{EB}$ = -(H$_{C(L)}$ + H$_{C(R)}$)/2, where H$_{C(L)}$ and H$_{C(R)}$ are the left and right intercepts of the magnetization curve with the field axis. A large value of H$_{EB}$ $\sim$ 4.65 kOe is observed at 10 K in a moderate cooling field of 1 kOe. It is noteworthy that a clear shift in the zero-field cooled (ZFC) M-H, however much smaller in comparison with the FC M-H is observed for this material (Zero field cooled exchange bias field, H$_{ZEB}$ = 0.4 kOe at 4 K). 
\begin{figure}[h]
\includegraphics[width=1.0\columnwidth]{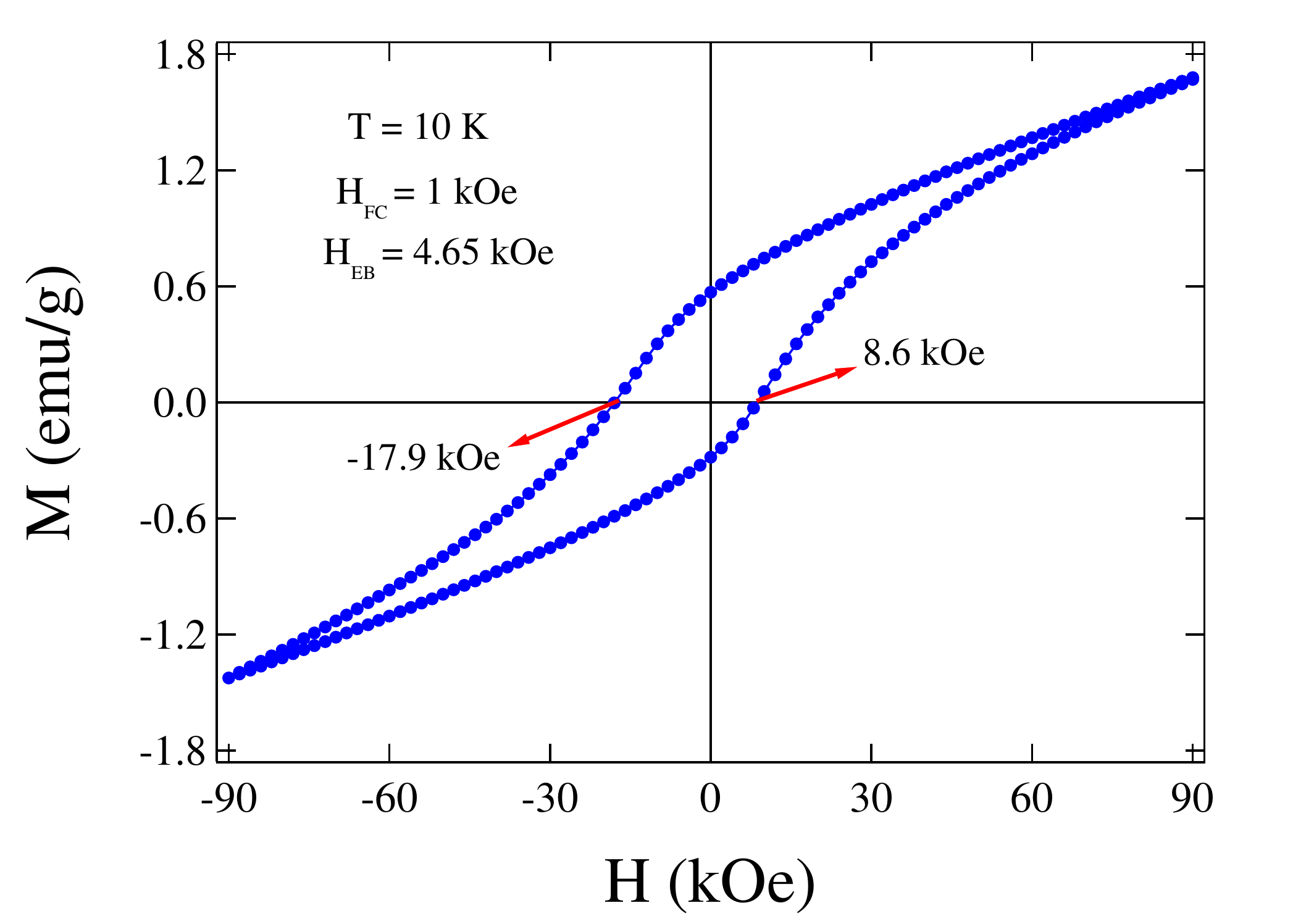}
\caption{\label{Fig3:dsk} Magnetic field dependent magnetization loop (M-H) for Sr$_{3}$YCo$_{4}$O$_{10.5}$ material at 10 K measured in a field cooling mode (FC, cooling field H$_{FC}$ = 1 kOe).}
\end{figure}
Recently, it has been shown that ferrimagnetism could isothermally induce the exchange anisotropy needed for the rarely observed ZFC EB (for instance in ferrimagnetically ordered Mn$_{2}$PtGa \cite{21} ZFC EB field = 1.75 kOe at 4.2 K and in polycrystalline La$_{1.5}$Ca$_{0.5}$CoMnO$_{6}$ ceramics \cite{22} ZFC EB field = 0.3 kOe at 2 K). Therefore, the existence of ferrimagnetism could be the origin of the observed ZFC exchange bias for the present material. However, the existence of a small remnant magnetic field ($\sim$ 2 - 5 Oe) in the superconducting coil of the magnetometer may induce this exchange anisotropy. Further studies (such as in ultra low magnetic field) are required to clarify the evolution of ZFC exchange bias. For the present study, we have focused on the moderate magnetic field (cooling field) induced exchange bias effect for the Sr$_{3}$YCo$_{4}$O$_{10.5}$ sample. Indeed, the moderate magnetic field induced large EB effect is of interest to fabricate the EB devices as it eliminates the requirement of external high magnetic field to create the unidirectional anisotropy.

	To explore the temperature dependence of EB properties, we have performed the temperature dependence of exchange bias effect. The sample was field-cooled (FC) to the measuring temperatures in an applied field of 1 kOe. Once the measuring temperature was reached, the M-H loops were measured between $\pm$ 90 kOe. The temperature variation of the exchange bias field (H$_{EB}$) and coercivity (H$_{C}$) is shown in Fig. 4.
	 
\begin{figure}[h]
\includegraphics[width=1.0\columnwidth]{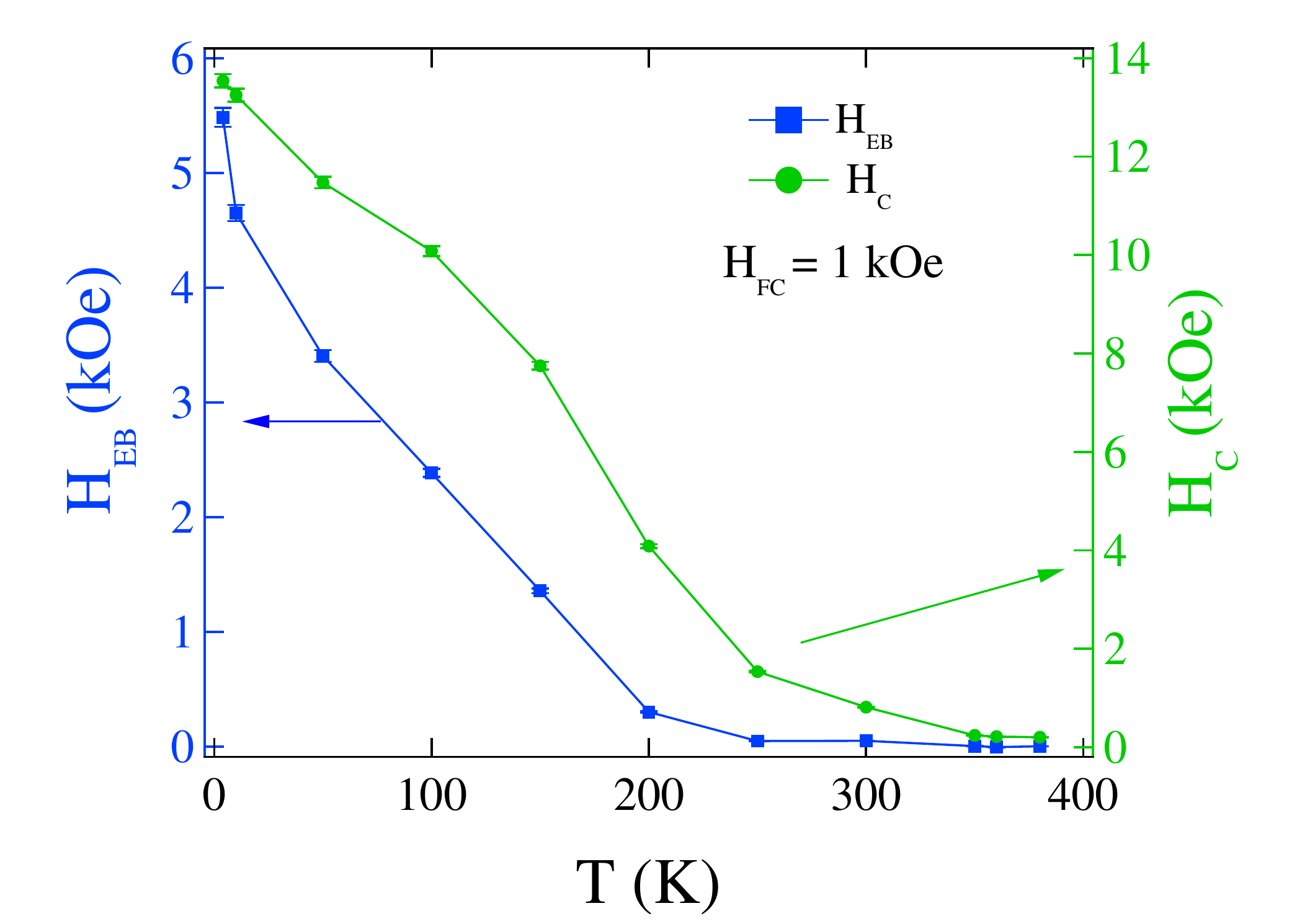}
\caption{\label{Fig4:dmn} Temperature dependence (left axis) of moderate field cooled (H$_{FC}$ = 1 kOe) exchange bias (H$_{EB}$) and (right axis) coercivity for the Sr$_{3}$YCo$_{4}$O$_{10.5}$ material.}
\end{figure}

The EB effect is started above room temperature (H$_{EB}$ = 50 Oe at 300 K) and with lowering the temperature H$_{EB}$ continuously increases and shows a maximum value of 5.5 kOe at 4 K. Remarkably, the value of exchange bias field for the present sample is extremely large in a moderate cooling field in comparison with material systems which are recognized for their ability to show giant exchange bias effect. For instance, at 4.2 K the state-of-the-art bulk material Mn$_{2.4}$Pt$_{0.6}$Ga \cite{13} shows H$_{EB}$ $\sim$ 1.5 kOe at a cooling field of 1 kOe, bulk Mn$_{50}$Ni$_{42}$Sn$_{8}$ \cite{23} Heusler alloy shows H$_{EB}$ $<$ 2.5 kOe at a cooling field of 1 kOe at 2 K. Ferrimagnetic double perovskite with composition Ba$_{2}$Fe$_{1.12}$Os$_{0.88}$O$_{6}$ shows H$_{EB}$ $\sim$ 4 kOe at a cooling field of 1kOe at 2 K \cite{14}. At the same time, the coercivity (Fig. 4) also persists up to the magnetic ordering temperature ($\sim$ 350 K) of the compound. 
  
   Along with the conventional horizontal shift (magnetic field axis), M - H loops are also shifted along the vertical direction (magnetization axis). This is likely due to some intrinsic magnetic moment which does not take place in the magnetization reversal for applied fields of magnitude smaller than 90 kOe. The disordered FM/AFM contributions in the sample could be the origin of the vertical shift of the M - H loops. Similar vertical shift of the hysteresis loops is also observed in several bulk samples \cite{13,14}. However, in order to calculate exchange bias fields from the real pinning effect, we have corrected our data by symmetrizing the M - H loops with respect to the magnetization axis. Fig. 5 shows the temperature variation of exchange bias fields from these symmetrized M - H loops. As expected, a large drop in the exchange bias shift is observed after symmetrizing the M - H loops. 
   \begin{figure}[h]
\includegraphics[width=1.0\columnwidth]{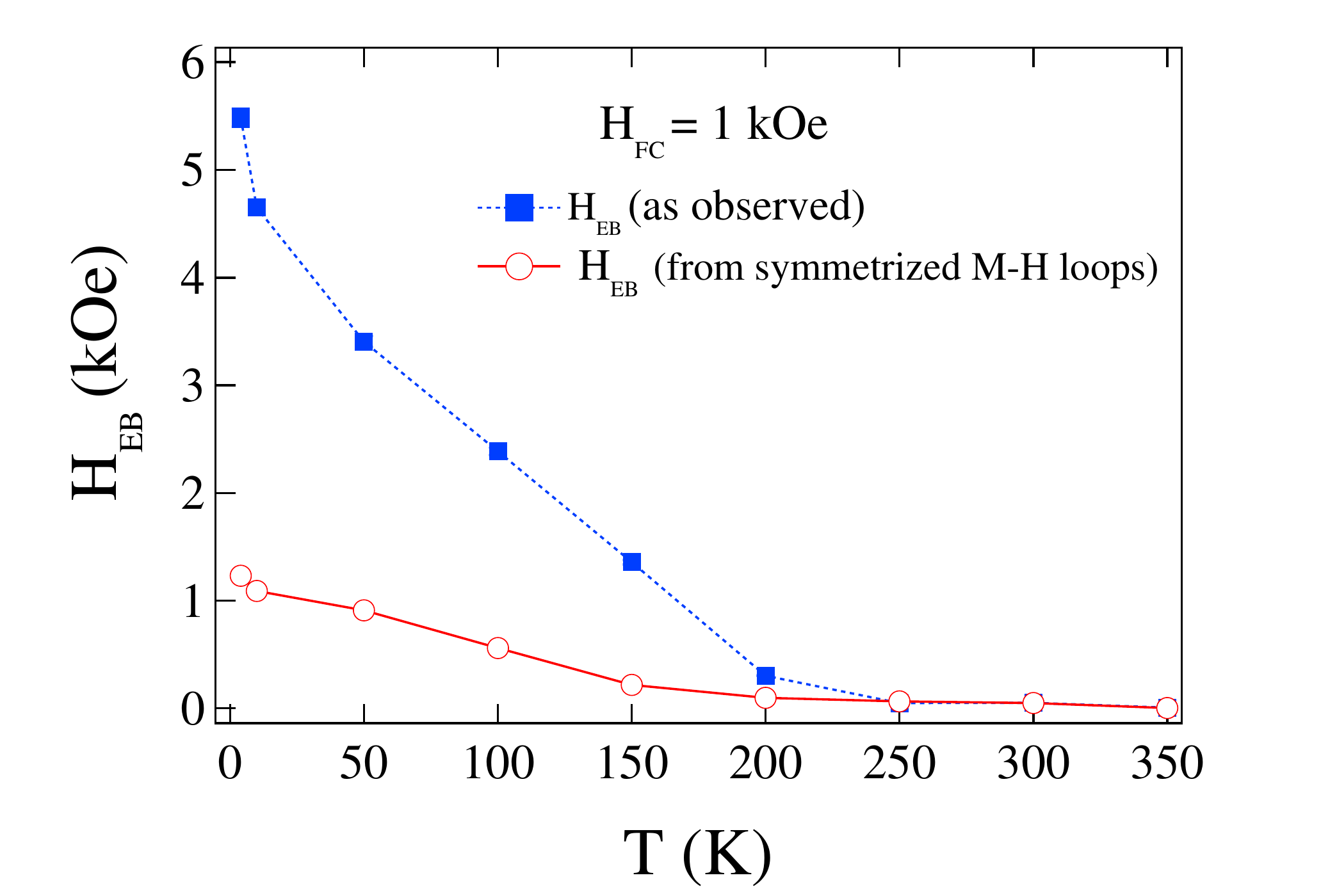}
\caption{\label{Fig5:dmn} Temperature dependence of exchange bias fields from symmetrized M - H loops for the Sr$_{3}$YCo$_{4}$O$_{10.5}$ material. In comparison with the as observed exchange bias field (blue and dotted line) the exchange bias shift is fairly reduced after symmetrizing the M - H loops.}
\end{figure}
   
	An important characteristic related to the EB property, training effect is measured for the present sample and shown in Fig. 6. Here the uncompensated spin configuration at the interface may relax due to the repetition of the M-H loop, and the H$_{EB}$ depends on the number of successive hysteresis loops measured. An enlarged view of the low field region of successive M-H loops at 4 K is shown in the inset of Fig. 6. It clearly shows that the exchange bias decreases monotonically as the cycle number (n) increases. A power law relationship is usually suggested to describe the number of field cycle (n) dependence of exchange bias \cite{24}.
\begin{equation}
H_{EB}(n)-H_{E\infty} = k/\sqrt{n}  ,
\label{eqn1:h}
\end{equation}
Where H$_{EB}$ (n) is the exchange field for the nth cycle. H$_{E\infty}$ is the exchange field for n = ${\infty}$, and k is a system dependent constant. The obtained value of H$_{E\infty}$ is 4.6 kOe, which will be the remnant H$_{EB}$ in the sample and the fitted data (solid line) is shown in Fig. 6. Furthermore, our experimental data matched well with the recursive formula suggested by Binek \cite{25} i.e., 
\begin{equation}
H_{EB}(n+1)-H_{EB} (n) = \gamma (H_{EB}(n)-H_{E\infty})^{3}  ,
\label{eqn2:hm}
\end{equation}                   																								                                                   
where $\gamma$ = 1/2k$^{2}$. It is noteworthy that Eq. 1 is only applicable for data starting from n = 2, while Eq. 2 describes also the apparent outlier for n = 1. 
\begin{figure}[h]
\includegraphics[width=1.0\columnwidth]{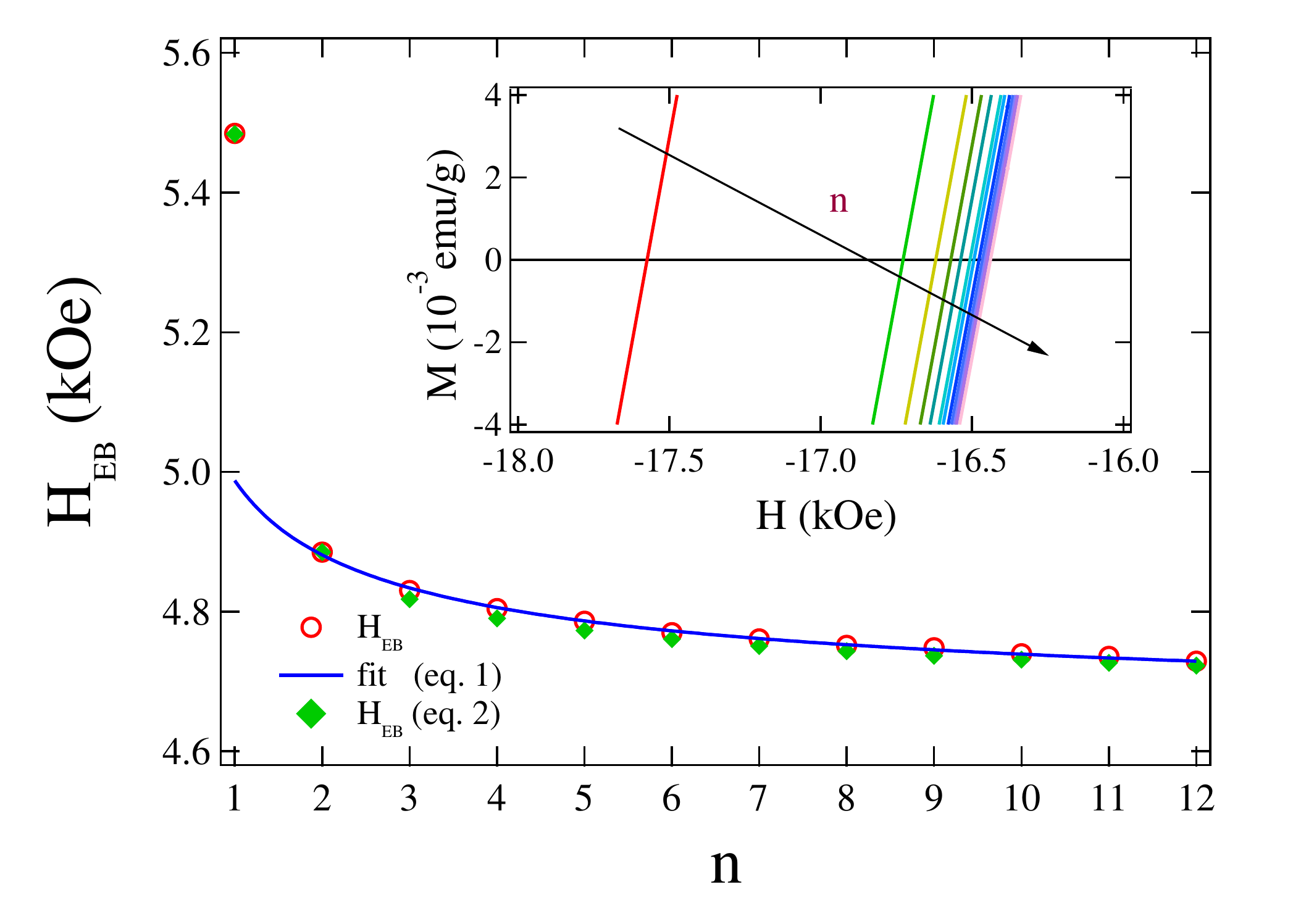}
\caption{\label{Fig5:dn} The exchange bias field (H$_{EB}$) vs number of field cycle (n) (red open circles) obtained from the training effect magnetic hysteresis loops (M-H) at 4 K. The inset shows an enlarged view of the consecutive M-H loops. Arrow indicates the direction of increase in field cycle (n).}
\end{figure}

	As described previously, recent advances in nearly compensated ferrimagnetic materials (for instance Mn$_{3-x}$Pt$_{x}$Ga\cite{13}) highlighted a novel approach in designing new materials with a large EB. The observed giant EB effects in those systems were attributed to the presence of ferrimagnetic clusters in the compensated host, which are formed as a consequence of antisite disorder. For the present layered Sr$_{3}$YCo$_{4}$O$_{10.5}$ sample, there are two different Co sites in the structure, one is oxygen-replete (Co2, see Fig. 1) and another is oxygen-deficient (Co1). The ferrimagnetic structure was found to form with a Co$^{3+}$ HS state and IS state ordering (AFM aligned) in the oxygen-replete layers, while the Co cations in the oxygen deficient site are nearly compensated \cite{18}. Therefore, similarly to the nearly compensated Heusler alloys Mn$_{3-x}$Pt$_{x}$Ga, the existence of ferrimagnetic clusters in the compensated host originates exchange bias effect in this system. The layered structure type of the present material with large exchange bias in a moderate cooling field and magnetic ordering above room temperature indicates that 314 - type cobaltates are a promising class of compounds for designing magnetic materials with huge exchange bias in low/moderate magnetic field. Appropriate chemical doping could result in a large EB at room temperature. For instance, Sr doping causes a significant enhancement in the exchange bias properties for La$_{2-x}$Sr$_{x}$CoMnO$_{6}$ \cite{26}.
\section{Conclusions}
In summary, we have shown moderate cooling field induced exchange bias for 314 - Sr$_{3}$YCo$_{4}$O$_{10.5}$ material. The exchange bias is observed below 345 K and reaches a maximum value of 5.5 kOe at 4 K. The large exchange anisotropy originates from the exchange interaction between the compensated host and ferrimagnetic clusters, which is intrinsic in this sample. We would like to emphasize that the Sr$_{3}$YCo$_{4}$O$_{10.5}$ sample shows a large value of exchange bias in a moderate cooling field in comparison with material systems which are recognized for their potentiality to show giant exchange bias. Indeed, the moderate magnetic field induced (cooling field) large EB would be interesting for the technological applications as it eliminates the requirement of the high external magnetic field to create the unidirectional anisotropy. Alongside the appearance of a sizable EB in a moderate magnetic field, the feasibility of room temperature application suggests exploring similar coherent materials that can be implemented in thin film form. At the same time, ZFC EB, however much smaller in comparison with the FC EB is observed for this material.  The intrinsic ferrimagnetism in the layered structure could isothermally induce the exchange anisotropy needed for the ZFC EB.\\

\ack
R. P. S.  acknowledges Science and Engineering Research Board (SERB), Government of India for the Ramanujan Fellowship through Grant No. SR/S2/RJN-83/2012. S. M acknowledges Science and Engineering Research Board, Government of India for the NPDF fellowship (Ref. No. PDF/2016/000348).

\end{document}